\documentclass[referee]{aa-mpa}

\usepackage{epsf}

\newcommand{\be}{\begin{displaymath}}
\newcommand{\ee}{\end{displaymath}}
\newcommand{\bea}{\begin{eqnarray}}
\newcommand{\eea}{\end{eqnarray}}

\begin{document}

\thesaurus{06(08.03.2; 08.05.3; 08.09.3; 08.12.1; 08.18.1)}

\title{Episodic lithium production by extra-mixing in red giants}

\author{P.A.~Denissenkov\inst{1,2}
        \and
        A.~Weiss\inst{2}}

\institute{Astronomical Institute of the St. Petersburg University,
           Bibliotechnaya Pl.~2, Petrodvorets, 198904~St.\,Petersburg,
           Russia
           \and
           Max-Planck-Institut f\"{u}r Astrophysik,
           Karl-Schwarzschild-Str.~1, 85740 Garching,
           Federal Republic of Germany}

\offprints{A.~Weiss (weiss@mpa-garching.mpg.de)}

\date{Received; accepted}

\titlerunning{Episodic lithium production in red giants}

\authorrunning{Denissenkov \& Weiss:}

\maketitle

\vskip 3.0cm
\begin{abstract}

The recent discovery of low-mass red giants with enhanced
atmospherical Li abundance in galactic low-metallicity stellar
clusters adds to the mysterious phenomenon of
the Li-rich giants. We propose a combined scenario for the Li-enrichment:
engulfing a giant planet (or brown dwarf) by a red giant (external source)
activates inside the giant the ``$^7$Be-mechanism'' producing Li
internally. This episodical Li-production can happen at any time
on the red giant branch and is naturally followed by Li-depletion
as is observed. Limitations of our scenario are discussed as well.


\vskip 2.0cm
\centerline{\em Accepted for publication in Astron.\ Astrophys.\ Letters}

\end{abstract}

\clearpage

\section{Introduction}

In low-mass red giants ($M\leq 2.5\,M_\odot$) the excursion of the
base of the convective envelope (BCE) into the radiative zone separating it
from the hydrogen-burning shell (HBS) initiates the first dredge-up
episode during which the surface composition experiences modest changes.
The best indicators of
the 1st dredge-up are the ratio $^{12}$C/$^{13}$C and the Li abundance,
both decreasing. A decrease of the C and an increase of N
are also predicted (\cite{bs99}).
In the overwhelming majority of the field red giants (K giants)
the surface Li abundance is low.
However, there are several percent of K giants which possess
surprisingly high Li abundances, sometimes exceeding the initial value
of $\log\varepsilon(^7{\mbox{Li}})\approx 3$ for Pop.~I stars
(Wallerstein \& Sneden 1982; Hanni 1984; Brown et al. 1989; Berdyugina \&
Savanov 1994; Da Silva et al. 1995; De la Reza et al. 1996, 1997).
It was not until recently when the evolutionary status of these Li-rich
giants (LIRGs) became clearer. Owing to the more accurate luminosities
estimated through the Hipparcos parallaxes Jasniewicz et al. (1999)
have concluded that most of the LIRGs are past the 1st dredge-up
red giant branch (RGB) stars (the low $^{12}$C/$^{13}$C ratios
observed in the LIRGs support this) and hence the initial abundance of
Li in their atmospheres could not have been preserved.

De la Reza et al. (1997), Jasniewicz (1999) and
other researchers argue convincingly for an internal mechanism of Li
production in LIRGs. 
In both papers quoted the ``$^7$Be-transport'' mechanism
(\cite{cf71}) is considered as the most promising
internal source of Li.
This internal mechanism has recently received new
support: Castilho et al. (1999) have reported that Be -- as predicted
-- is very depleted (about 10 times) in both LIRGs they studied.
In this {\em Letter} we propose a combined scenario of Li-enrichment:
a giant planet (or brown dwarf) is engulfed by a red giant; this external
source or event activates inside the giant the ``$^7$Be-mechanism''
producing then Li internally. Before presenting our model in Sect.~3,
we will discuss existing evidence for extra-mixing in red giants (Sect.~2),
because this is crucial for scrutinizing any scenario.
Sect.~4 concludes the paper.

\section{Extra-mixing in red giants}

The 1st dredge-up episode ends when the BCE stops its excursion into
the star and begins to retreat. According to the standard theory, after
this moment the surface composition is not expected to change further on the RGB.
However, this expectation is not supported by observations. It appears
that in low-mass field red giants the surface abundances of C, N, and Li
as well as the $^{12}$C/$^{13}$C ratio, and in globular cluster red giants
even O, Na, Mg and Al, do continue to change
up to the RGB tip. This discrepancy has been quite satisfactorily explained
by red giants models with extra-mixing placed between the HBS and the BCE
(see \cite{dt00}, and references therein).
Such extra-mixing is commonly believed to work only
in radiative regions with a nearly uniform composition. Therefore, it is
expected to come into play after the HBS, moving outwards in mass,
will have erased the H-discontinuity left behind by the retreating BCE.
Standard model calculations show that the HBS arrives at the H-discontinuity
while the star is still on the RGB only
in low-mass stars, which is well confirmed by observations
(\cite{cc94}).

\begin{figure}
\epsfxsize=10cm
\centerline{\epsffile [30 220 560 720] {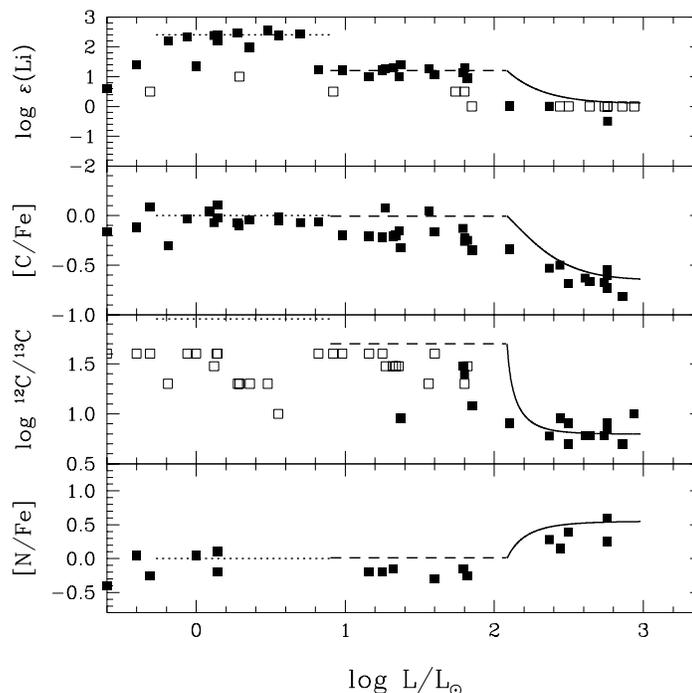}}

\caption[]{Li, C, N abundances and $^{12}$C/$^{13}$C ratios in field stars
           with accurate luminosity estimates in the metallicity range
           $-2\leq$ [Fe/H] $\leq -1$ (Gratton et al.~1999); open squares
           are upper (for Li) or lower (for $^{12}$C/$^{13}$C) limits).
           Dotted and dashed horizontal segments mark the main sequence and 1st
           dredge-up luminosity ranges, respectively. Solid lines were
           calculated with extra-mixing modeled by diffusion with
           depth $\delta m_{\mbox{\tiny mix}}=0.12$ and
           rate $D_{\mbox{\tiny mix}}=5\cdot 10^8$~cm$^2\cdot$s$^{-1}$.}

\label{gratton}

\end{figure}

Quite recently Gratton et al. (1999) have determined Li, C, N, O and Na
abundances and $^{12}$C/$^{13}$C ratios for a large sample of field stars
with accurate luminosity estimates in the metallicity range
$-2 \leq$ [Fe/H] $\leq -1$. In Fig.~\ref{gratton} we compare
$\log\varepsilon(^7\mbox{Li})$, [C/Fe] ([A/B] means
$\log[n(\mbox{A})/n(\mbox{B})]_{\mbox{\small star}} -
\log[n(\mbox{A})/n(\mbox{B})]_\odot$), $\log\,^{12}\mbox{C}/^{13}\mbox{C}$ and
[N/Fe] for these stars with results of calculations obtained
with the method and code of Denissenkov \& Weiss (1996) in which extra-mixing is
modeled by diffusion in a post-processing approach, which uses full
stellar evolution models as background models for the parameterized diffusion and
nucleosynthesis. One can see that the behaviour of
the plotted abundances on the upper-RGB ($\log L/L_\odot > 2$,
following the nomenclature of Gratton et al.\ 1999)
is quite well reproduced by
the diffusive mixing with a depth $\delta m_{\mbox{\small mix}} = 0.12$
($\delta m$ is a relative mass coordinate such that $\delta m = 0$
at the bottom of the HBS and $\delta m = 1$ at the BCE) and a rate
$D_{\mbox{\small mix}} = 5\cdot 10^8$~cm$^2\cdot$s$^{-1}$ (for details
about method and results, see also \cite{dw96}).

From the upper panel of Fig.~\ref{gratton} one can infer that {\em (i)}
most of the Pop.~II main sequence stars preserve the initial Li abundance in their atmospheres
($\log\varepsilon(^7{\mbox{Li}})\approx 2.3$ for Pop.~II stars),
{\em (ii)} during the 1st dredge-up
Li is diluted exactly down to the level predicted by the standard
theory (\cite{sb99}), and {\em (iii)} extra-mixing on the upper-RGB further decreases
the surface Li abundance; extra-mixing is therefore a necessary
ingredient to explain this behaviour.

\begin{figure}
\epsfxsize=10cm
\centerline{\epsffile [30 220 560 720] {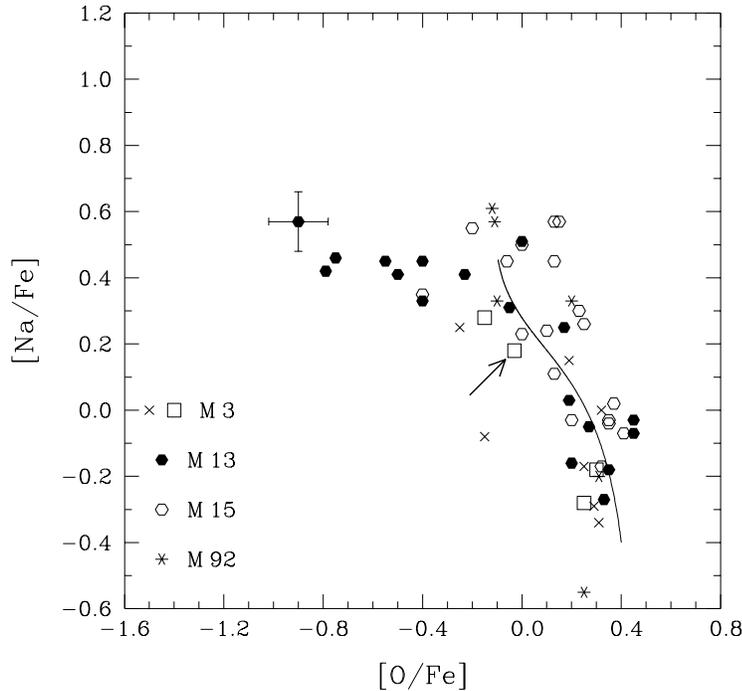}}

\caption[]{The global [Na/Fe] vs. [O/Fe] anticorrelation in globular cluster
           red giants and its theoretical reproduction by diffusive
           extra-mixing with depth $\delta m_{\mbox{\tiny mix}}=0.06$ and
           rate $D_{\mbox{\tiny mix}}=5\cdot 10^8$~cm$^2\cdot$s$^{-1}$ (solid line).
           Open squares are M\,3 giants from the recent Li-study of
           Kraft et al. (1999). The Li-rich giant IV-101 is shown by
           the arrow. Data from Kraft et al.\ (1992; M3), Kraft et
           al.\ (1997; M13), Sneden et al.\ (1997; M15), Shetrone
           (1996; M92)}
\label{ona}

\end{figure}

Contrary to the field Pop.~II giants which show neither O depletion nor Na
enhancement, in globular clusters there are star-to-star variations of
both O and Na on the RGB. Even more important is the fact that
in globular cluster red giants Na anticorrelates with O (Fig.~\ref{ona},
symbols). A summary of the observational status can be found in
Sneden (1999). The global anticorrelation of [Na/Fe] vs. [O/Fe]
can be explained by extra-mixing as well (\cite{dw96}), but in this case deeper mixing
is required. In Fig.~\ref{ona} we compare observational data with a
sample calculation (solid line, calculated with $\delta m_{\mbox{\small mix}} = 0.06$,
$D_{\mbox{\small mix}} = 5\cdot 10^8$~cm$^2\cdot$s$^{-1}$).
The corresponding evolution of the surface Li abundance for this case
is plotted in Fig.~\ref{lilgl} (line 3a).
We note that in all our cases (Fig.~\ref{gratton}, solid lines;
Figs.~\ref{ona} and \ref{lilgl}) the calculations with extra-mixing start from
the same red giant model with $M=0.8\,M_\odot$, $\log L/L_\odot=2.1$ and
a heavy elements content of $Z=5\cdot 10^{-4}$ ([Fe/H] $\approx\log Z/Z_\odot=-1.58$).
Because of the shallower mixing in the models of
the field Pop.~II stars no Na was produced in that case, as observed
(\cite{grattonea99}).

To conclude, extra-mixing (with specific values for mixing depth and
speed) is necessary to explain abundance trends in the {\em majority}
of field giants and abundance anomalies in a {\em large} number of
globular cluster giants. We now turn to the even more peculiar effect
of Li-richness.

\section{A solution to the problem of Li-rich giants}

Following their first discovery that the ``$^7$Be-mechanism'' can naturally
work in luminous intermediate-mass asymptotic giant branch (AGB) stars
(\cite{sb92}), Sackmann \& Boothroyd (1999) have demonstrated that under certain
conditions the same process can produce Li on the first giant branch,
too. In AGB stars, $^7$Be freshly minted in the reaction
$^3$He($\alpha,\gamma)^7$Be is quickly mixed away to a cooler region
(where Li produced in the reaction $^7$Be(e$^-,\nu)^7$Li can survive)
by ordinary convection whereas in RGB stars
some extra-mixing is required for this.

The majority of field LIRGs have
circumstellar dust shells (\cite{dlrea96}) and a large
number of additional LIRGs have been discovered among
stars with IR excess. This feature seems to be the only one to distinguish
the LIRGs from ordinary K giants and led De la Reza et al. (1996)
to propose a scenario linking the high Li abundances in these stars to
the evolution of circumstellar shells. In this scenario {\em every} low-mass
red giant passes through a short phase during which some internal mechanism
initiates atmospherical Li enrichment accompanied by a prompt mass-loss
event. De la Reza et al. (1996) have calculated evolutionary paths
(in the IRAS color-color diagram) of the detached shells and inferred that
the whole cycle completes in about $10^5$ years, the very fast
initial increase of the surface Li abundance (during the first several thousand
years) being followed by the much longer period (up to $10^5$ years) of
Li depletion.


Recently, Siess \& Livio (1999) have considered an original external scenario:
a red giant engulfs an orbiting body of sub-stellar mass (brown dwarf
or giant planet) which has the initial abundance of Li left unprocessed.
This body deposits its Li into the giant's envelope and also causes
a shell ejection as a consequence of associated processes (mass accretion
near the BCE where the body is expected to dissolve and subsequent thermal
expansion of the overlying layers; for details see the cited paper).
This scenario has an obvious disadvantage: it cannot account for Li
abundances exceeding the initial one.

\begin{figure}
\epsfxsize=10cm
\centerline{\epsffile [30 220 560 720] {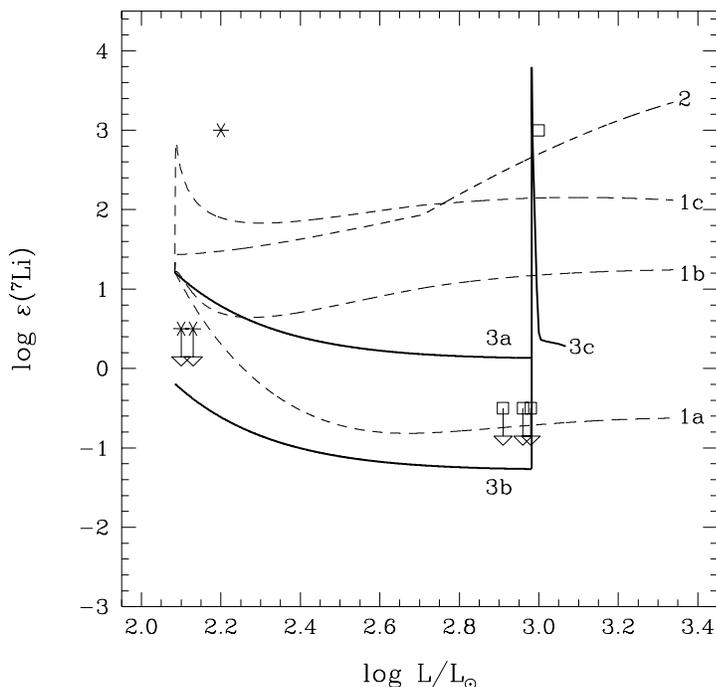}}

\caption[]{Evolution of the surface $^7$Li abundance on the upper-RGB caused by extra-mixing.
           The mixing parameters used in our calculations (for which we used
           code and method described in Denissenkov \& Weiss 1996) are
           {\bf 1}: $\Delta\log T=\log T(\delta m=0)-\log T(\delta m_{\mbox{\tiny mix}})=0.26$
           ($\delta m_{\mbox{\tiny mix}}\geq 0.14$),
           {\bf a} -- $D_{\mbox{\tiny mix}}=10^9$~cm$^2\cdot$s$^{-1}$,
           {\bf b} -- $D_{\mbox{\tiny mix}}=5\cdot 10^9$~cm$^2\cdot$s$^{-1}$,
           {\bf c} -- $D_{\mbox{\tiny mix}}=10^{11}$~cm$^2\cdot$s$^{-1}$;
           {\bf 2}: $\Delta\log T=0.36\ (\delta m_{\mbox{\tiny mix}}\geq 0.17)$,
           $D_{\mbox{\tiny mix}}=10^{11}$~cm$^2\cdot$s$^{-1}$;
           {\bf 3a,b}: $\delta m_{\mbox{\tiny mix}}=0.06$,
           $D_{\mbox{\tiny mix}}=5\cdot 10^8$~cm$^2\cdot$s$^{-1}$;
           {\bf 3c}:\,$\delta m_{\mbox{\tiny mix}}$ decreases with time by $0.01$ every
           $8\cdot 10^3$~years from 0.16 to 0.06 and after that retains the constant
           value 0.06, $D_{\mbox{\tiny mix}}=10^{12}$~cm$^2\cdot$s$^{-1}$ during
           all this time. Open squares and asterisks are M3 and Berkeley~21 giants
           from the recent Li-studies of Kraft et al. (1999) and Hill \&
           Pasquini (1999), respectively.}

\label{lilgl}

\end{figure}

In this {\em Letter} we propose a combined scenario in which engulfing of a giant
planet by a red giant initiates the internal ``$^7$Be-mechanism'':
It takes into account results of quite recent publications where
for the first time extremely high Li abundances have been measured
in {\em cluster} giants. These are the stars IV-101 ([Fe/H] $=-1.50$) in the globular cluster
M\,3 (\cite{kraftea99}) and T33 ([Fe/H] $=-0.58$) in the metal-poor open cluster
Berkeley~21 (\cite{hp99}). In both cases a Li abundance of
$\log\varepsilon(^7{\mbox{Li}})\approx 3.0$ has been reported.
The LIRGs IV-101 and T33 are plotted in Fig.~\ref{lilgl} in comparison with
5 Li-normal giants from the same studies. One realizes that
an episodical Li-enrichment can happen at any time on the upper-RGB,
independent of the red giant's evolutionary state, thus
indicating an external source.
Fig.~\ref{ona} supports this conclusion: Both the Li-rich giant IV-101
(open square and arrow) and another Li-normal one (open square) close
to it have Na increased and O decreased and fit well to the global
[Na/Fe] vs. [O/Fe] anticorrelation.
At the same time extra-mixing
with the parameters adjusted to reproduce this anticorrelation
(Fig.~\ref{ona}, solid line) fails to make LIRGs (Fig.~\ref{lilgl},
lines 3a and 3b for two different values of initial Li). It appears
that after having been exposed for a rather long
time ($\sim 3\cdot 10^7$ years) to the ``ordinary'' extra-mixing which
is responsible to the Na-O-anomalies, the star IV-101 -- but not the
other one --
experienced something which suddenly changed its extra-mixing parameters
to values appropriate for Li-production.

From our model calculations we found that
the ``$^7$Be-mechanism'' can efficiently synthesize Li and after that maintain
its high abundance for a long time only if
$0.12\leq\delta m_{\mbox{\small mix}}\leq 0.18$ and, even more important,
only if $D_{\mbox{\small mix}}\geq 10^{11}$~cm$^2\cdot$s$^{-1}$
(Fig.~\ref{lilgl}, lines 1c and 2). Denissenkov \& Tout (2000)
have proposed Zahn's rotation-driven meridional
circulation and turbulent diffusion (\cite{zahn92}; \cite{mz98}) as
a physical mechanism for extra-mixing in low-mass red giants.
It turns out, however, that with Zahn's mechanism, values of
$D_{\mbox{\small mix}}\geq 10^{11}$~cm$^2\cdot$s$^{-1}$ can be obtained only
as upper limits for rotation close to the Keplerian one.
In a scenario with engulfing a planet such fast rotation is explained
naturally as a result of transferring the planet's orbital
angular momentum to the giant's envelope (\cite{sl99}). The next
question then is
how to get the correct mixing depth in this scenario.

The dashed lines in Fig.~\ref{lilgl} are similar to those shown in Fig.~10 of
Sackmann \& Boothroyd (1999). They are the result of calculations
under the assumption
that mixing depth and rate favourable for the Li-production are
constant on the upper-RGB. One of them (like our line 2)
has even been used to interpret a LIRG near the RGB tip in the globular
cluster NGC\,362 by Smith et al. (1999). However, such a straightforward
interpretation is not so simple because: {\em (i)} mixing under these
conditions does not
produce Na nor deplete O as is observed in IV-101; {\em (ii)} it explains
neither the Li-depletion immediately following the Li-production nor
the rather short time-scale for the whole cycle;
{\em (iii)} it requires a very unusual,
precise and long-term tuning of the mixing parameters; the tuning appears
to be unusual because it assumes shallow but extremely fast mixing compared
to that reproducing the [Na/Fe] vs. [O/Fe] anticorrelation; it would
be more natural
to expect that faster mixing should be deeper as well.

Thus we propose the following explanation of how the correct mixing depth
could appear in the engulfing scenario. According to Siess \& Livio (1999)
the giant planet (or brown dwarf) dissolves near the BCE in a red giant.
After that the rotation profile in the radiative zone takes a step-like
shape with a steep increase of the angular velocity up to about
a local Keplerian value at the point of deepest penetration by the planet.
In the course of the subsequent evolution the HBS moves outwards in mass and
after $\sim 8\cdot 10^5$ years will reach the step in
the rotation profile. During a time interval of $\sim 8\cdot 10^4$ years
this step will be crossing a zone $0.06\leq\delta m\leq 0.16$ where and when
the Li-production becomes efficient. Thus we do not need to fix the mixing depth
to a preferred value. Instead, the natural growth of the helium core
assures that suitable depths will be encountered and very fast mixing
(due to the planet's engulfing) produces Li during this passage
(Fig.~\ref{lilgl}, line 3c).

\section{Conclusion}

A great advantage of the proposed solution is
that it can account not only for the Li-production but also for the subsequent
Li-depletion. Indeed, we find that after the mixing depth has reduced
to less
than $\sim 0.08$, Li begins to be destroyed on a time-scale 
consistent with the results of De la Reza et al. (1996).
It should be emphasized that it is even more difficult to deplete Li
quickly after its production than to produce it, and our scenario deals with
this naturally.

At the same time, however, it cannot be applied to the Li-rich
($\log\varepsilon(^7{\mbox{Li}})\approx 1.8$) star V42 in M5
(\cite{cfg98}), which appears to be a (low-mass) post-AGB star. Due to
the timescales, the Li-enrichment cannot have happened already during
the first red giant phase. On the other side, since V42 is only as
bright as the RGB-tip ($M_{\rm Bol} = -3.38$), but hotter and thus
smaller, capturing a companion would have happened already on the
RGB. Thus, our scenario fails for this star, which otherwise appears
to be typical M5 member, showing standard $\alpha$-enhancement and even
the O-Na-anticorrelation (\cite{cfg98}). We can only speculate that
its Li-overabundance happened (via the Cameron-Fowler mechanism)
during the AGB, where additional deep mixing initiated hot bottom
burning as is standard in intermediate-mass stars (\cite{sb92}). Due
to the thin envelope of this star very modest extra mixing might
already be sufficient. We will investigate this possibility in
forthcoming work.

Our scenario does neither provide a direct link to the dust
shell formation. Siess \& Livio (1999) have ascribed the shell detachment
to an increased mass loss during the planet's engulfing but in our scenario
this event is separated from the Li-enrichment episode by a time interval of
$\sim 7\cdot 10^5$ years. The following two speculations towards a solution of
this problem can be envisaged: (1) The mass of the radiative zone is negligible,
and the angular velocity inside it scales as $\propto r^{-2}$ (\cite{dt00});
hence, the ratio of the centrifugal acceleration to the gravity scales
as $\propto r^{-1}$; as after engulfing the planet
this ratio is expected to become close to unity near the BCE, then during the subsequent
inward excursion of the step in the rotation profile
it will surely exceed unity somewhere in the radiative zone, which may
initiate dynamical processes of the angular momentum transfer outwards;
the latter may be responsible for the increased mass loss. (2) The process of
planet engulfing itself may be associated with various dynamical and
thermodynamical processes, for instance, a deepening of the convective
envelope (\cite{sl99}), which may redistribute
the material with the high angular momentum
throughout the radiative zone;
in this case the fast mixing will be able to penetrate the zone of
correct mixing depths from the very beginning. Whether such phenomena
happen in a real red giant can be verified only by 3D hydrodynamical
simulations.






\begin{acknowledgements}

We wish to express our gratitude to the referee, R.~Kraft, who pointed
out the existence of V42. 
P.A.D. acknowledges the warm hospitality of the staff of
the Max-Planck-Institut f\"{u}r Astrophysik where this study was
carried out.

\end{acknowledgements}

{}

\end{document}